\documentclass[aps,prl,showpacs,twocolumn,superscriptaddress]{revtex4}

\usepackage{amssymb}
\usepackage{epsfig}
\usepackage{graphicx}
\usepackage{amsmath}
\usepackage{array,color}
\usepackage{float}

\begin{document}

\title{Dirac monopoles with polar-core vortex induced by
spin-orbit coupling in spinor Bose-Einstein condensates}
\author{Ji Li}
\affiliation{Institute of Physics, Chinese Academy of Sciences,
Beijing 100190, China}
\author{Yan-Mei Yu}
\affiliation{Institute of Physics, Chinese Academy of Sciences,
Beijing 100190, China}
\author{Lin Zhuang}
\affiliation{School of Physics, Sun Yat-Sen University,
Guangzhou 510275, P. R. China}
\author{Wu-Ming Liu}
\email{wliu@iphy.ac.cn}\affiliation{Institute of Physics,
Chinese Academy of Sciences, Beijing 100190, China}

\begin{abstract}
  We report Dirac monopoles with polar-core vortex induced by spin-orbit
  coupling in ferromagnetic Bose-Einstein condensates,
  which are attached to two nodal vortex lines along the vertical axis.
  These monopoles are more stable in the time scale of experiment
  and can be detected through directly imaging vortex lines.
  When the strength of spin-orbit coupling increases, Dirac monopoles
  with vortex can be transformed into those with square lattice.
  In the presence of spin-orbit coupling, increasing the strength of
  interaction can induce a cyclic phase transition from Dirac monopoles
  with polar-core vortex to those with Mermin-Ho vortex. The spin-orbit
  coupled Bose-Einstein condensates not only provide a new unique platform
  for investigating exotic monopoles and relevant phase transitions,
  but also can preserve stable monopoles after a quadrupole field is turned off.

\end{abstract}
 \maketitle

Dirac monopoles \cite{Dirac} have attracted more and more attention
in a wide area of research including solid state physics \cite{Castelnovo,Morris,Pollard,Zhou,Khomskii,Bovo},
the quantum field \cite{Ray,Ray2}, and other systems \cite{Cardoso,Goddard,Brekke,Bakker}.
In particular, the recent realization of spinor Bose-Einstein condensates (BECs), due to
many possible order-parameter manifolds,
provides an ideal platform for creating monopoles \cite{Martikainen,Pietila} and
others topological nontrivial structures \cite{Williams,Hall,Khawaja,Choi}. So far,
both the monopole with one terminating nodal line \cite{Ray} and
the isolated monopole without such nodal line \cite{Ray2} have been realized in
the BECs. Theoretically,
several types of monopoles have
been investigated \cite{Busch,Garc,Stoof,Savage,Ruokokoski,Pietil,Conduit,Cho,Solnyshkov,Kiffner},
including two-dimensional monopoles \cite{Busch,Garc}, the monopoles in
antiferromagnetic system \cite{Stoof}, and two-component monopoles \cite{Pietil}.
A majority of studies on monopoles in spinor
BECs have been only limited to the systems
with spin-dependent interaction. However, spin-orbit (SO) coupling, the interaction between the spin of a
quantum particle and its momentum, has not been considered.

The SO coupling in the BECs can be controlled and tuned by
using optical fields or a sequence of pulsed homogeneous magnetic fields
\cite{Lin,Rus,Liu1,Zhang1,Ji,Campbell,Lan,Anderson1,Anderson,Wang1,Cheuk},
which provides opportunities to search for
novel quantum states in BECs \cite{Wang,Su,Liu,Xu,Sinha,Hu,Gopalakrishnan,Li,Han}.
These novel quantum states are
based on the fact that the SO coupling makes the internal states coupled to their
momenta. Meanwhile, due to the SO coupling, the atoms with pseudo-spins are not
constrained by fundamental symmetries such as global symmetry and mirror symmetry. This will give
rise to the remarkable phenomena not encountered anywhere else in physics.
An immediate question is, whether the SO coupling induces unknown types of monopoles
that do not have an analogy in the case of spinor BECs without the SO coupling.

In this Letter, we find a new type of monopoles, the
Dirac monopoles with the polar-core vortex (M-PCV), induced by the SO coupling in
ferromagnetic BECs. Different from the case without the SO coupling \cite{Ray,Pietila},
here the monopoles
locates at the endpoints of two nodal lines along the vertical axis. Compared with previous work \cite{Ray,Ruokokoski},
in this work, the Dirac strings are not observed to split until $22.8$ ms,
indicating that M-PCV are more stable
and long-lived, which makes the potential experimental observation of
such the monopoles easier to be realized. We further demonstrate that
the monopoles with the square lattice (M-SL) occur under the strong SO coupling, which can be observed in a wide region of parameters.
We find, for the first time, increasing the strength of spin-independent
interaction can induce a cyclic phase transition from M-PVC to those with Mermin-Ho vortex (M-MHV) in the presence
of weak SO coupling.

\begin{figure}
\includegraphics[width= 0.48\textwidth]{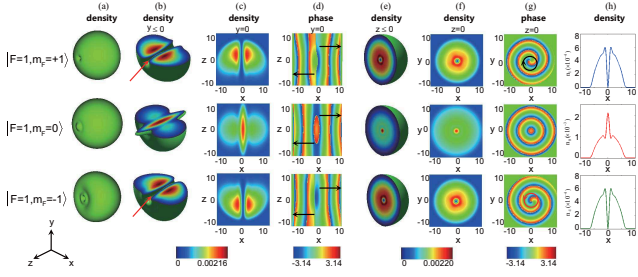}
\caption{(Color online). (a) Isosurface of particle densities. (b) Particle densities for $y\leq0$, where the red arrows indicate the location of two nodal lines (Dirac strings). The phenomenon for $x\leq0$ is the same that for $y\leq0$. (c) and (d) Densities and phase distributions of the $y=0$ planes, in (d), the phases at both sides of the $y=0$ planes are inverse, which is denoted by the black arrows. (e) Particle densities for $z\leq0$. (f) and (g) Densities and phase distributions of the $z=0$ planes, the singly vortex is denoted by the black circle of arrow and the singly antivortex is denoted by the red circle of arrow. (h) The one-dimensional density distributions the corresponding (f). Here,  the dimensionless SO coupling strength $\kappa=2$, the dimensionless spin-dependent interaction parameter $\lambda_{2}=-75$, the dimensionless spin-independent interaction parameter $\lambda_{0}=7500$, the dimensionless strength of the magnetic field gradient $B_{1}=0.6$, and the isotropic optical trap $\omega_{r}=\omega_{z}=2\pi\times250$Hz \cite{Pietila}.} \label{fig1}
\end{figure}

\begin{figure}
\includegraphics[width= 0.48\textwidth]{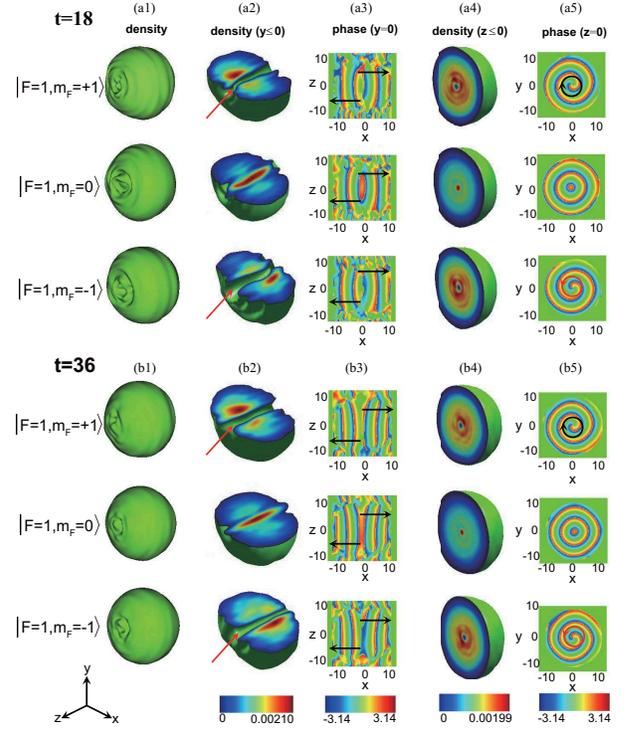}
\caption{(Color online). Real time evolutions of M-PCV. (a1)-(a5) $t=18$. (b1)-(b5) $t=36$. (a1) and (b1) Isosurface of the particle densities. (a2) and (b2) Particle densities for $y\leq0$, where the red arrows indicate the location of the nodal lines
(Dirac strings). (a3) and (b3) Phase distributions of the $y=0$ planes. Note that the phases at both sides of the $y=0$ planes are inverse, which is denoted by the black arrows. (a4) and (b4) Particle densities for $z\leq0$. (a5) and (b5) Phase distributions of the $z=0$ planes, the singly vortex is denoted by the black circle of arrow and the singly antivortex is denoted by the red circle of arrow. Here, the real time $t'=0.64t$ msec and other parameters are
same as in Fig.1.
} \label{fig2}
\end{figure}

We consider the monopoles that arise from the three-dimensional
spin-1 BECs with a two-dimensional SO
coupling \cite{Anderson} and a controllable magnetic field \cite{Ray}.
In the mean-field approximation, the Hamiltonian for the spin-1
BECs in an optical trap is written as \cite{Pietila,Savage,Ruokokoski,Wang,Pu,Su,Liu}
\begin{eqnarray}
\begin{split}
H\!=\!&\int d^3\mathbf{r}\bigg\{\mathbf{\Psi}^\dagger\left(T\!+\!V_{opt}(\mathbf{r})
\!+\!\mathcal{V}_{so}
\!+\!g_{F}\mu_{B}\mathbf{B}(\mathbf{r})\cdot\mathcal{F}\right)\mathbf{\Psi} \\
&\!+\!\left(\frac{c_{0}}{2}n^{2}\!+\!\frac{c_{2}}{2}[(n_{1}\!-\!n_{-1})^{2}
\!+\!2|\Psi^{\ast}_{1}\Psi_{0}\!+\!\Psi^{\ast}_{0}\Psi_{-1}|^{2}]\right)\bigg\},
\end{split}
\end{eqnarray}
where $\mathbf{\Psi} =[\Psi_{1}(\mathbf{r}),\Psi_{0}(\mathbf{r}),\Psi_{-1}(\mathbf{r})]^{T}$
is the order parameter of the BECs with normalization
$\int d^{3}\mathbf{r}\mathbf{\Psi}^{\dagger}\mathbf{\Psi}=N$, and $N$ is the
total particle number. The kinetic energy $T=-\hbar^{2}\nabla^{2}/(2m)$.
The total particle density is defined by $n=\sum_{m}n_{m}$,
wherein $n_{m}=|\Psi_{m}(\mathbf{r})|^{2}$ with
$m=0, \pm1$. The optical trapping potential $V_{opt}(\mathbf{r})=m[\omega^{2}_{r}(x^{2}+y^{2})+\omega^{2}_{z}z^{2}]/2$,
where $\omega_r$ and $\omega_z$ are the radial and axial trapping frequencies,
and $m$ is the mass of a $^{87}$Rb atom.
The vector of spin-1 matrices is defined by $\mathcal{F}=(\mathcal{F}_{x},\mathcal{F}_{y},\mathcal{F}_{z})^{T}$,
wherein $\mathcal{F}_{x}$, $\mathcal{F}_{y}$ and $\mathcal{F}_{z}$ are the $3\times3$ Pauli spin-1 matrices.
The SO coupling term is written as $\mathcal{V}_{so} = -i\hbar(\kappa_{x}\mathcal{F}_{x}\partial_{x}+\kappa_{y}\mathcal{F}_{y}\partial_{y})$,
where $\kappa_{x}$ and $\kappa_{y}$ are the strengths of the SO coupling.
We define $\kappa_{x}=\kappa_{y}=\kappa$ for isotropic
SO coupling (Rashba-type) and $\kappa_{x}\neq\kappa_{y}$ for anisotropic SO coupling. The external magnetic field is written as
$\mathbf{B}(\mathbf{r})=B_{1}^{'}(x\widehat{e}_{x}+y\widehat{e}_{y})+B_{2}^{'}z\widehat{e}_{z}$,
where the condition $2B_{1}^{'}+B_{2}^{'}=0$ must be satisfied according
to Maxwell's equation $\nabla\cdot\mathbf{B}=0$. The Land$\acute{e}$
factor $g_{F}=-1/2$ and $\mu_{B}$ is
the Bohr magnetion. For the interaction terms, the coupling parameters
are given by $c_{0} = 4\pi\hbar^{2}(a_{0}+2a_{2})/3m$ and $c_{2} = 4\pi\hbar^{2}(a_{2}-a_{0})/3m$,
where $\hbar$ is the Planck constant and $a_{0,2}$ are two-body s-wave
scattering lengths for total spin $0,2$. We choose $a_{2}=(100.4\pm0.1)a_{B}$ for
total spin channel $F_{total}=2$ and $a_{0}=(101.8\pm0.2)a_{B}$ for total
spin channel $F_{total}=0$ \cite{Stenger,van,Stamper-Kurn}, where $a_{B}$ is
the Bohr radius. The time, the energy, the strength
of the SO coupling, and the strength of the magnetic field gradient
are scaled by $\omega^{-1}$, $\hbar\omega$, $\sqrt{\hbar\omega/m}$,
and $\omega\hbar/(g_{F}\mu_{B}a_{h})$, respectively. The stationary
states of the monopoles are obtained (see Supplemental Material \cite{sm}) by using the standard
imaginary-time propagation combined with finite-difference methods \cite{Dalfovo,Zhang,Bao}. The dynamic
evolutions of the monopoles are obtained using the split-operator
combined with the Crank-Nicolson method, the time step of dynamic simulation is $10^{-4}/\omega$.

\begin{figure}
\includegraphics[width= 0.48 \textwidth]{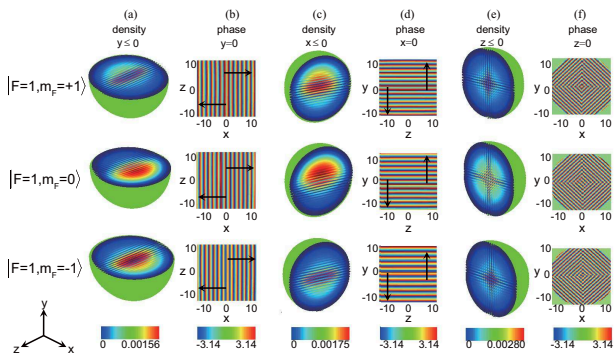}
\caption{(Color online). (a) Particle densities for $y\leq0$. (b) Phase distributions of the $y=0$ planes, the phases at both sides of the $y=0$ planes are inverse, which is denoted by the black arrows. (c) Particle densities for $x\leq0$. (d) Phase distributions of the $x=0$ planes, the phases at both sides of the $x=0$ planes are also inverse, which is denoted by the black arrows. (e) Particle densities for $z\leq0$. Note that the density distribution in the $x$-$y$ plane ($z=0$) represents square lattice structure. (f) Phase distributions of the $z=0$ planes. Here, $\kappa=16$ and other parameters are
same as in Fig.1.} \label{fig3}
\end{figure}

We first study
the structures of the monopoles without the SO coupling.
A doubly quantized vortex line splits into two singly quantized vortex
lines partly in the $m_{F}=+1$ and $-1$ components, because here
a homogeneous bias field is not considered in the absence of the SO coupling.
The phases of the vortex line along the $z$ and
$y$ axes are opposed, which is similar to the vortex-antivortex pair
\cite{Devreese,Chibotaru}. In fact, the doubly quantized vortex will
split into two singly quantized vortices absolutely as the time goes
on \cite{Shin,Huhtam}. Therefore, we can suppose that the monopole
is metastable (M-MS) in the absence of the SO coupling \cite{sm}.
Next, we start to study how the SO coupling gives rise to exotic monopole
structures. When the SO coupling is weak, the M-PCV are found. The structures of the M-PCV
represent a singly vortex line in the $m_{F}=+1$ component, a soliton in
the $m_{F}=0$ component and a singly antivortex line in the $m_{F}=-1$
component (Fig. 1).
Compared with the monopoles in BECs without
the SO coupling \cite{Ray,Pietila}, in the present system, there
exists two monopoles located at the endpoints of two nodal lines
along the vertical axis [as highlighted by the
red arrows in Fig. 1(b)], which is believed to be caused by the
interaction of the spin of a particle with its motion. Meanwhile,
being different from the monopoles in spin ices \cite{Castelnovo,Morris,Pollard,Zhou,Khomskii,Bovo},
our results demonstrate the fundamental quantum features and topological
structures of the monopoles predicted by Dirac \cite{Dirac}. A topological
defect has a longer lifetime, which is beneficial for its experimental
observation. Therefore, we perform the dynamic simulations for
the M-PCV. The simulations show that
the structures of the monopoles keep the original shape (Fig. 1) from $t=18$
to $t=36$ [Figs. 2(a1)-2(b5)]. Especially, the nodal lines are not expanded
and still exist in the condensates [Figs. 2(a2) and 2(b2)]. Furthermore,
during the dynamic evolution, as seen in the phase profile of the
wave function in the $z=0$ plane, the singly vorticity is well maintained
and no vortex splitting is observed [Figs. 2(a5) and 2(b5)].
Compared with the dynamics of the monopoles in the absence of the SO
coupling \cite{Ray}, where the doubly quantized vortex splits into two
singly vortices after roughly $10ms$. In our case, the quantized vortex
is not observed to split until $t=36$ ($22.8ms$), confirming that the M-PCV
are more stable in the time scale
of experiment. Therefore, we can expect that
the experimental observation of such monopoles should be
more practical in the case of the SO coupling. In addition, the dynamics of the monopoles are also
stable after the magnetic field is turned off, which suggests that the SO
coupling can preserve stable monopoles \cite{sm}.

\begin{figure}
\includegraphics[width= 0.48\textwidth]{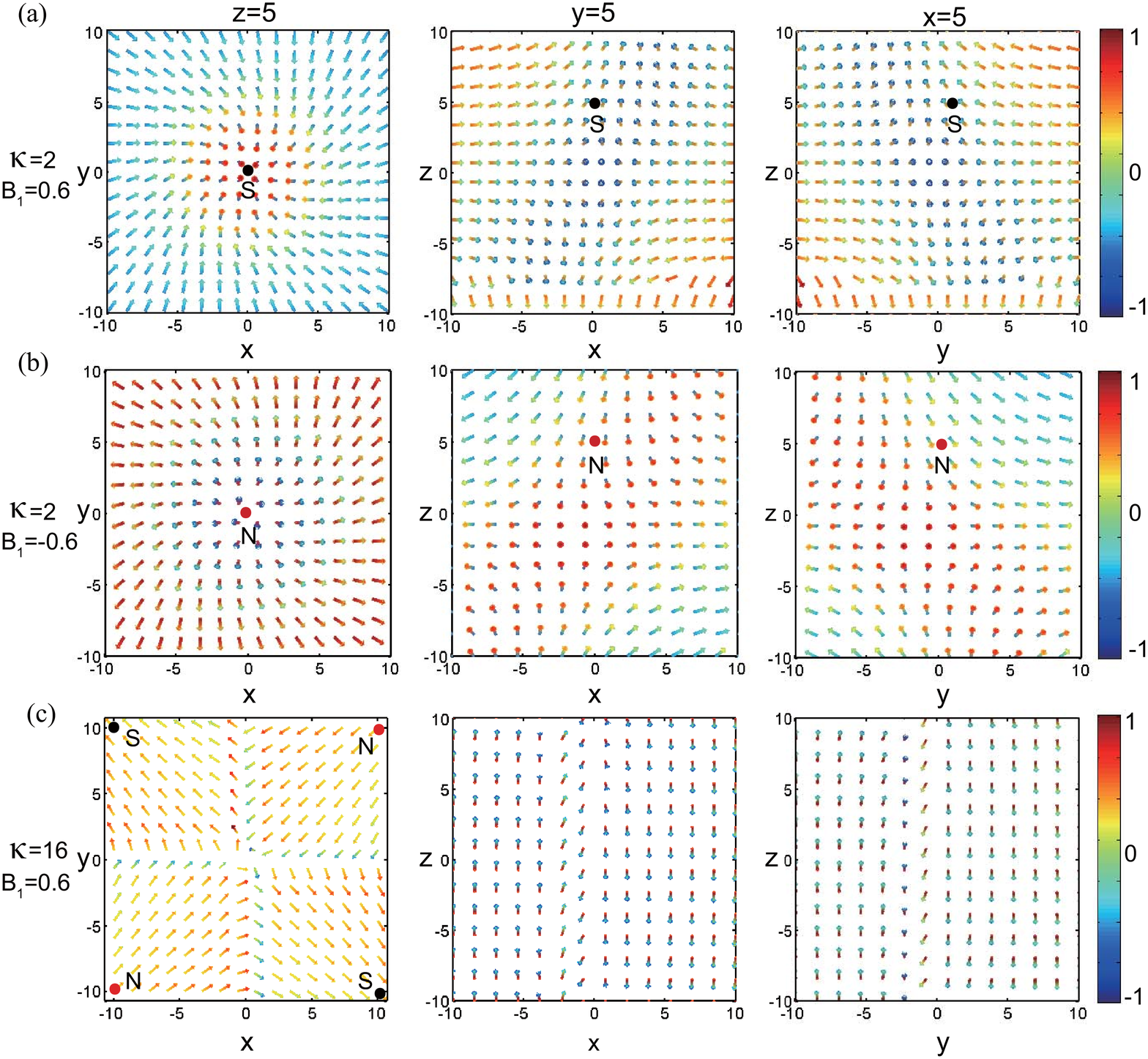}
\caption{(Color online). (a) Spin textures of M-PCV describe the south magnetic poles (the black dot S). (b) Spin textures of the antimonopoles describe the north magnetic poles (the red dot N).
(c) Spin textures of M-SL represent spin configurations of the south magnetic poles (the black dot S) and north magnetic poles (the red dot N). Other parameters are same as in Fig.1.} \label{fig4}
\end{figure}

\begin{figure}
\includegraphics[width= 0.48\textwidth]{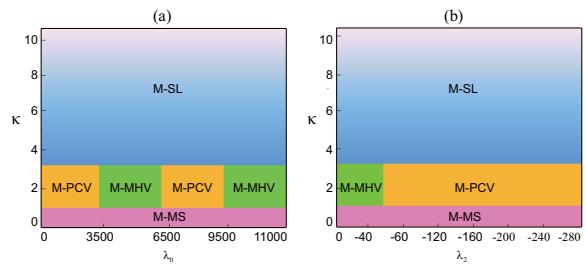}
\caption{(Color online). (a) Phase diagram of the monopole with $\kappa$ and $\lambda_{0}$ for a given $\lambda_{2}$.
(b) Phase diagram of the monopole with $\kappa$ and $\lambda_{2}$ for a given $\lambda_{0}$.} \label{fig5}
\end{figure}

We also demonstrate that the M-PCV can occur in the presence
of the oblate trap \cite{sm}. The influence of the magnetic field gradient on the
monopoles is also investigated \cite{sm}.
In order to identify the effect of the interaction on the monopoles,
we also perform the simulations that decrease $\lambda_{2}$ for a given $\kappa$. The simulations show that
the M-MHV is obtained, which represents a soliton in the $m_{F}=+1$ component,
a singly vortex line in the $m_{F}=0$ component and a doubly quantized
vortex line in the $m_{F}=-1$ component \cite{Mizushima}. This result shows that such the monopole
structure can exist in the condensates not only without SO coupling \cite{Ray,Pietila},
but also with SO coupling \cite{sm}. When the SO coupling strength increases,
the M-SL are found, as shown in Figs. 3(a)-3(f).
The square lattice in the central zone is very distinct, as compared with that
in the surrounding zone. Meanwhile, the density distribution in the longitudinal
direction is of the stripe structure.
More importantly, the M-SL are not affected by the
interactions, because the lowest-energy
state is not affected by the interactions for the strong SO coupling. Meanwhile, the
antimonopoles emerge in the system, which suggests that the periodic magnetic
monopole can occur in the presence of strong SO coupling. Finally, we also consider an
anisotropic SO coupling case, which shows that the monopole vanishes. Because the Dirac string can be removed in the presence of asymmetric SO coupling \cite{sm}.

The spinor BECs can be considered as a magnetic system,
which reflects physical properties of topological defects \cite{Stamper-Kurn}.
We therefore study topological spin textures of the monopoles. the components of the spin
vector are given by $F_{x}=\frac{1}{\sqrt{2}}[\psi_{1}^{\ast}\psi_{0}+\psi_{0}^{\ast}(\psi_{1}+\psi_{-1})+\psi_{-1}^{\ast}\psi_{0}]$,
$F_{y}=\frac{i}{\sqrt{2}}[-\psi_{1}^{\ast}\psi_{0}+\psi_{0}^{\ast}(\psi_{1}-\psi_{-1})+\psi_{-1}^{\ast}\psi_{0}]$, and
$F_{z}=|\psi_{1}|^{2}-|\psi_{-1}|^{2}$ \cite{Su,Liu,Mizushima,Kasamatsu,Kawakami,Mizushima2,Kasamatsu2}. The spin textures of the M-PCV are shown in Fig. 4(a).
The spin aligns with the radially inward hedgehog distribution in the $x$-$y$ plane,
representing the spin texture of a south magnetic pole, while the spin textures
in the $x$-$z$ and $y$-$z$ planes show the cross hyperbolic distribution \cite{Mizushima2,Kita}.
It has been shown that in the
absence of the SO coupling, the spin texture shows the north magnetic pole
\cite{Ruokokoski}. In our case, the SO coupling changes the spin direction,
which leads to the fact that the north magnetic pole can be transformed into
the south magnetic pole. As shown in Fig. 4(b), the spin textures of the antimonopoles
are observed, which represents the north magnetic
poles. For the case of the strong SO coupling,
as shown in Fig. 4(c), the spin textures are divided into four portions in the
$x$-$y$ plane. The spin texture behaves as the ferromagnetic distribution at
each portion. The spin orientations at two diagonal portions are opposite,
which reflects the structures of monopoles and antimonopoles that locate in
the boundaries of the condensates. The spin textures in the $x$-$z$ and $y$-$z$
planes form the spin stripe. In addition, the spin texture in the case of the anisotropic SO
coupling is also investigated \cite{sm}.
Finally, we further prove that the M-PCV are stable through the dynamic evolution
of the spin texture \cite{sm}.

We now consider the interplay between the SO coupling
and the interaction, due to the coupling of the spin of a particle
with its motion induced by magnetic pulses and the transformation
of magnetic order induced by the interaction, which leads to the
rich phase diagrams of the monopoles. For a given $\lambda_{2}$ being -75, the phase diagram as
functions of $\kappa$ and $\lambda_{0}$ is shown in Fig. 5(a). The monopole
is metastable when $\kappa$ is less than a critical value $\kappa_{c}=0.8$.
For $0.8<\kappa\leq3$, the M-PCV exist
when $0<\lambda_{0}\leq3000$ and $5000<\lambda_{0}\leq8000$, while the
M-MHV emerge when $3000<\lambda_{0}\leq5000$
and $8000<\lambda_{0}\leq11000$. This indicates that a cyclic phase
transition from the M-PCV to the M-MHV can occur with the increasing $\lambda_{0}$. Note
that if there is no the SO coupling, such the cyclic phase transition
will not occur. The M-SL occur when
$\kappa>3$, confirming that at strong SO coupling the monopoles
are not affected by the interactions and can exist in the wide
parameter region. Furthermore, the phase diagram as functions of
$\kappa$ and $\lambda_{2}$ for a given $\lambda_{0}$ being 7500 is
shown in Fig. 5(b). For $0.8<\kappa\leq3$,
there exists the M-MHV when $\lambda_{2}$
is less than 60, while the M-PCV emerge
when $\lambda_{2}$ is more than 60. This result suggests that the
increasing the spin-dependent interaction only leads to the direct
phase transition from the M-MHV to
M-PCV.

We finally briefly discuss the experimental feasibility of creating the monopoles in SO coupled BECs (see Supplemental Material \cite{sm}).
We consider spin-1 BECs of alkali $^{87}$Rb atoms, where particle number $N\approx0.6(1.8)\times10^{5}$.
First of all, a quadrupole field is applied and turned on, which is realized by a pair of Helmholtz coils.
Then, when a point source of the superfluid flow is formed \cite{Pietila}, the quadrupole field is turned off. The pulsed magnetic fields are turned on, which creates two-dimensional SO coupling \cite{Anderson}.  We take some parameters from the recent experiments \cite{Ray,Pietila,Anderson}, which includes the isotropic optical trap $\omega_{r}=\omega_{z}=2\pi\times250$Hz, the anisotropic optical trap $\omega_{r}\approx2\pi\times160$ Hz and $\omega_{z}\approx2\pi\times220$ Hz, the constant bias magnetic field $B^{(0)}=20$G, and the quadrupole field gradient $B_{1}^{'}=0.03\sim0.1$ T/m. For the M-PCV, the SO coupling strength $\kappa\sim 0.8-3$. We find the dynamic
period of the monopoles $t'\sim11ms$ , which is much shorter than the lifetime of the BECs and can be observed stably in experiments.

In conclusion, we have shown that the weak SO coupling leads to the
emergence of the M-PCV that
are stable and long-lived,
and the strong SO coupling leads to the emergence of the M-SL in spinor BECs. We have predicted the rich phase diagrams of the
monopoles by changing the SO coupling strength, the spin-dependent
interaction, and the spin-independent interaction. Such monopoles can be proved by
means of imaging the vortex lines in a real experiment.
This work paves the way for
future explorations of the monopole with respect to gauge field,
topological defects and the corresponding dynamical stability in quantum systems.

\begin{acknowledgments}
 This work was supported by the NKRDP under grants Nos. 2016YFA0301500, 2012CB821305,
 NSFC under grants Nos. 11434015, 61227902, 61378017, 61376014,
 SKLQOQOD under grants No. KF201403,
 SPRPCAS under grants No. XDB01020300, XDB21030300.
\end{acknowledgments}

\begin{widetext}

\section{Supplementary Material}

In this supplementary material, we present the details on calculation of the stationary states with respect to the monopoles,
experimental setup of creating the monopoles in spin-orbit (SO) coupled Bose-Einstein condensates (BECs),
the structures of the monopoles in the absence of the SO coupling, the monopoles with the Mermin-Ho vortex (M-MHV), the effect of the
anisotropic optical trap on the monopoles, the effect of the quadrupole field gradient on the monopoles,
ground states for the anisotropic SO coupling, spin textures for the anisotropic SO coupling,
the dynamic evolution of the monopoles in the absence of the quadrupole field, and the dynamic evolution of the spin texture.

\section*{Calculation of the stationary states with respect to the monopoles}

We investigate the stationary states of the monopoles in the SO coupled spinor BECs ,
which is based on the Gross-Pitaevskii
mean-field theory. The wave functions of spin-1 BECs are formulated as the dimensionless coupled Gross-Pitaevskii equations \cite{1,2,3,4,7,5,6}

\begin{eqnarray}\label{GP1}
\begin{split}
i\frac{\partial\psi_{1}}{\partial t}=&(-\frac{1}{2}\nabla^{2}+V+\lambda_{0}\rho+\lambda_{2}(\rho_{1}+\rho_{0}-\rho_{-1})+B_{2}z)\psi_{1} \\
&+B_{1}(x-iy)\psi_{0}+\kappa(-i\partial_{x}-\partial_{y})\psi_{0}+\lambda_{2}\psi^{\ast}_{-1}\psi_{0}^{2},
\end{split}
\end{eqnarray}
\begin{eqnarray}\label{GP2}
\begin{split}
i\frac{\partial\psi_{0}}{\partial t}=&(-\frac{1}{2}\nabla^{2}+V+\lambda_{0}\rho+\lambda_{2}(\rho_{1}+\rho_{-1}))\psi_{0} \\
&+B_{1}(x+iy)\psi_{1} +B_{1}(x-iy)\psi_{-1}+\kappa(-i\partial_{x}+\partial_{y})\psi_{1}\\
&+\kappa(-i\partial_{x}-\partial_{y})\psi_{-1}+
2\lambda_{2}\psi_{1}\psi_{-1}\psi_{0}^{\ast},
\end{split}
\end{eqnarray}
\begin{eqnarray}\label{GP3}
\begin{split}
i\frac{\partial\psi_{-1}}{\partial t}=&(-\frac{1}{2}\nabla^{2}+V+\lambda_{0}\rho+\lambda_{2}(\rho_{0}+\rho_{-1}-\rho_{1})-B_{2}z)\psi_{-1} \\
&+B_{1}(x+iy)\psi_{0} +\kappa(-i\partial_{x}+\partial_{y})\psi_{0}+
\lambda_{2}\psi_{1}^{\ast}\psi_{0}^{2},
\end{split}
\end{eqnarray}
where the dimensionless wave function $\psi_{j}=N^{-1/2}a_{h}^{3/2}\Psi_{j}$ $(j=1,0,-1)$ and
the total condensate density $\rho=\rho_{1}+\rho_{0}+\rho_{-1}$ with $\rho_{j}=|\psi_{j}|^{2}$ $(j=1,0,-1)$.
The dimensionless optical trapping potential $V(\mathbf{r})=\frac{1}{2}(\gamma_{r}^{2}x^{2}+\gamma_{r}^{2}y^{2}+
\gamma_{z}^{2}z^{2})$ with $\gamma_{r}=\omega_{r}/\omega$, $\gamma_{z}=\omega_{z}/\omega$ and
$\omega=min\{\omega_{r},\omega_{z}\}$. The dimensionless interaction strengths
$\lambda_{0}=4\pi N(a_{0}+2a_{2})/3a_{h}$ and $\lambda_{2}=4\pi N(a_{2}-a_{0})/3a_{h}$,
where $\hbar$ is the Planck constant and $a_{0,2}$ are two-body s-wave scattering
lengths for total spin $0,2$. The characteristic length of the harmonic trap is
defined as $a_{h}=\sqrt{\hbar/m\omega}$. The dimensionless strength of the magnetic
field gradient complies the condition $2B_{1}+B_{2}=0$. The time, the energy, the
strength of the SO coupling, and the strength of the magnetic field gradient are
scaled by $\omega^{-1}$, $\hbar\omega$, $\sqrt{\hbar\omega/m}$, and $\omega\hbar/(g_{F}\mu_{B}a_{h})$, respectively.

The stationary state wave functions of the monopoles are obtained
by using the standard imaginary-time propagation, which is combined with the second-order
centered finite-difference discretization and the backward/forward
Euler methods. By applying the transformation relation $t\rightarrow\tau=-\mathrm{i}t$,
the imaginary time evolution equations are expressed as follows:
\begin{eqnarray}
\begin{split}
\frac{\partial\psi_{1}}{\partial t}=&(\frac{1}{2}\nabla^{2}-V-\lambda_{0}\rho-\lambda_{2}(\rho_{1}+\rho_{0}-\rho_{-1})-B_{2}z)\psi_{1} \\
&-B_{1}(x-iy)\psi_{0}-\kappa(-i\partial_{x}-\partial_{y})\psi_{0}-\lambda_{2}\psi^{\ast}_{-1}\psi_{0}^{2},
\end{split}
\end{eqnarray}
\begin{eqnarray}
\begin{split}
\frac{\partial\psi_{0}}{\partial t}=&(\frac{1}{2}\nabla^{2}-V-\lambda_{0}\rho-\lambda_{2}(\rho_{1}+\rho_{-1}))\psi_{0} \\
&-B_{1}(x+iy)\psi_{1} -B_{1}(x-iy)\psi_{-1}-\kappa(-i\partial_{x}+\partial_{y})\psi_{1}\\
&-\kappa(-i\partial_{x}-\partial_{y})\psi_{-1}-
2\lambda_{2}\psi_{1}\psi_{-1}\psi_{0}^{\ast},
\end{split}
\end{eqnarray}
\begin{eqnarray}
\begin{split}
\frac{\partial\psi_{-1}}{\partial t}=&(\frac{1}{2}\nabla^{2}-V-\lambda_{0}\rho-\lambda_{2}(\rho_{0}+\rho_{-1}-\rho_{1})+B_{2}z)\psi_{-1} \\
&-B_{1}(x+iy)\psi_{0} -\kappa(-i\partial_{x}+\partial_{y})\psi_{0}-
\lambda_{2}\psi_{1}^{\ast}\psi_{0}^{2},
\end{split}
\end{eqnarray}
in addtion, the average energy of the system is expressed as follows:
\begin{eqnarray}
\begin{split}
E(\psi_{1},\psi_{0},\psi_{-1})= &\int_{\Omega}\bigg\{\sum_{m=0,\pm1}\psi_{m}^{\ast}h_{d}\psi_{m}+\frac{\lambda_{0}}{2}\rho^{2}+\frac{\lambda_{2}}{2}(\rho_{1}+\rho_{0}-\rho_{-1})\rho_{1} \\
&+\frac{\lambda_{2}}{2}(\rho_{1}+\rho_{-1})\rho_{0}+\frac{\lambda_{2}}{2}(\rho_{0}+\rho_{-1}-\rho_{1})\rho_{-1} \\
&+\lambda_{2}(\psi_{-1}^{\ast}\psi_{0}^{2}\psi_{1}^{\ast}+\psi_{-1}{\psi_{0}^{\ast}}^{2}\psi_{1})+B_{2}z(\rho_{1}-\rho_{-1})\\
&+B_{1}(x-\mathrm{i}y)(\psi_{1}^{\ast}\psi_{0}+\psi_{0}^{\ast}\psi_{-1})+B_{1}(x+\mathrm{i}y)(\psi_{0}^{\ast}\psi_{1}+\psi_{-1}^{\ast}\psi_{0}) \\
&+\kappa[\psi_{1}^{\ast}(-\mathrm{i}\partial_{x}-\partial_{y})\psi_{0}+\psi_{0}^{\ast}(-\mathrm{i}\partial_{x}+\partial_{y})\psi_{1}\\
&+\psi_{0}^{\ast}(-\mathrm{i}\partial_{x}-\partial_{y})\psi_{-1}+\psi_{-1}^{\ast}(-\mathrm{i}\partial_{x}+\partial_{y})\psi_{0}]\bigg\}d\Omega.
\end{split}
\end{eqnarray}
In order to solve equations (6)-(8), we use second-order
centered finite-difference for the spatial discretization and
the backward/forward Euler scheme to solve the corresponding
linear/nonlinear terms for the time discretization.
The periodic boundary condition is considered, the
computational grids are  120$\times$120$\times$120 that
corresponds to the volume 20$\times$20$\times$20 in the
units of $a_{h}^{3}$, and
the scale for three dimension system
is $34.2\mu m\times34.2\mu m\times34.2\mu m$. A initial trial state of the normalized
random number of complex-values is given. The final steady states
are not depend on the initial trial wave function. The average energy
decays monotonically with time until the steady states are reached.

\section*{Experimental setup of creating the monopoles in spin-orbit coupled Bose-Einstein condensates}
The experimental setup is shown schematically in Fig. 6(a).
We consider spin-1 BECs of alkali $^{87}$Rb atoms,
where the $^{87}$Rb BECs contain about $N\approx0.6(1.8)\times10^{5}$ atoms.
First of all, a quadrupole field is applied and turned on, which is realized by a pair of Helmholtz coils.
The strength of the magnetic field is zero at the center of the quandrupole field,
which corresponds to a point source of the superfluid flow. The superfluid
flow of the spinor condensates can be characterized by its vorticity $\Omega_{s}=\hbar\hat{\mathbf{e}}_{r'}/(mr'^{2})$,
where $r'$ is the sphere radius. The vorticity $\Omega_{s}$ is equivalent to the magnetic field
of a monopole that distributes radially outward in a hedgehog form. A monopole can be considered
as a point source of the superfluid flow \cite{1}. Then, when a point source of
the superfluid flow is formed, the quadrupole field is turned off. The pulsed magnetic
fields are turned on, which creates two-dimensional SO coupling \cite{8}.
The cloud of atoms is situated 50$\mu m$ above the surface of the atom chip.
A constant bias magnetic field $B^{(0)}\mathbf{e}_{z}$ is applied out of plane,
which leads to splitting of the magnetic sublevels. Two pairs of microwires
parallel to $\mathbf{e}_{x}$ and $\mathbf{e}_{y}$ provide the rf magnetic
fields $\mathbf{B}_{x}(\mathbf{r},t)$ and $\mathbf{B}_{y}(\mathbf{r},t)$.
In the first stage $(0\leq t<\tau)$, the rf magnetic field $\mathbf{B}_{x}(\mathbf{r},t)$
with the frequency $\omega_{1}$ leads to an effective coupling vector in the $x$
direction and a spin-dependent phase gradient in the $y$ direction, where the
SO coupling is written as $-i\hbar\kappa_{x}\mathcal{F}_{x}\partial_{x}$. In
the second stage $(\tau\leq t<2\tau)$, the rf magnetic field $\mathbf{B}_{y}(\mathbf{r},t)$
with frequency $\omega_{2}$ leads to an effective coupling vector in the $y$ direction
and a spin-dependent phase gradient in the $x$ direction, where the SO coupling
is written as $-i\hbar\kappa_{y}\mathcal{F}_{y}\partial_{y}$. When the rf magnetic
fields both $\mathbf{B}_{x}(\mathbf{r},t)$ and $\mathbf{B}_{y}(\mathbf{r},t)$ are
applied, the two-dimensional SO coupling is created, which is written as
$\mathcal{V}_{so}=-i\hbar(\kappa_{x}\mathcal{F}_{x}\partial_{x}+\kappa_{y}\mathcal{F}_{y}\partial_{y})$
in the first-order approximation for a sufficiently short duration $\tau$.
The strengths of the SO coupling $\kappa_{x}$ and $\kappa_{y}$ are determined
by the strengths of the magnetic field gradient of $\mathbf{B}_{x}(\mathbf{r},t)$
and $\mathbf{B}_{y}(\mathbf{r},t)$. Due to the SO coupling, the spin degeneracy
of three-component bosons is lifted and the free particle energy spectrum splits
into three energy branches with different helicity [Fig. 6(b)]. The Rashba ring
can be seen from the minimum energy spectrum, which is denoted by the black
circular ring in Fig. 6(b). We take some parameters from the recent experiments \cite{9,1,8},
which includes the isotropic optical trap $\omega_{r}=\omega_{z}=2\pi\times250$Hz,
the anisotropic optical trap $\omega_{r}\approx2\pi\times160$ Hz and
$\omega_{z}\approx2\pi\times220$ Hz, the constant bias magnetic field
$B^{(0)}=20$G, and the quadrupole field gradient $B_{1}^{'}=0.03\sim0.1$ T/m.

\begin{figure}
\includegraphics[width= 0.76\textwidth]{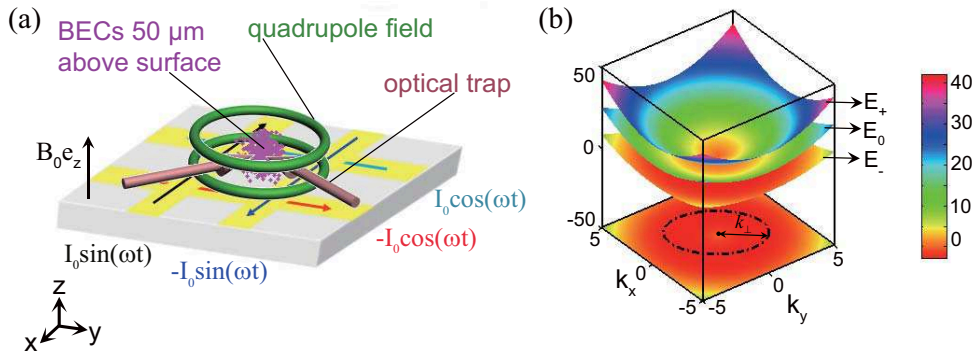}
\caption{(Color online).  Experimental setup of
creating the monopoles in SO coupled
BECs. (a) Experimental geometry of creating the quadrupole
field and the SO coupling. A pair of Helmholtz
coils are used for creating the quadrupole fields,
where the pink particles represent $^{87}$Rb atoms
and the kermesinus arrows show beam paths of the
optical trap. The pulsed magnetic field are used
for creating the SO coupling, where the cloud of atoms
is located 50$\mu m$ above the surface of an atom chip.
A bias field $B_{0}$ along the $z$ axis induces the magnetic
splitting. Two pairs of parallel microwires are embedded
in the atom chip, which produces the amplitude modulated
rf field along $x$ and $y$ axis, respectively, these two
effective coupling vectors in $x$-$y$ plane induce effective
two-dimensional SO coupling in the first-order approximation to short enough duration.
(b) Energy spectrum. The single-particle energy spectrum of
the monopole with SO coupling strength $\kappa=2$ in $k_{x}-k_{y}$
plane is divided into three branches $E_{+}$, $E_{0}$ and $E_{-}$,
the black circular ring represents the Rashba ring from the minimum energy spectrum.} \label{figs1}
\end{figure}

\begin{figure}
\includegraphics[width= 0.48\textwidth]{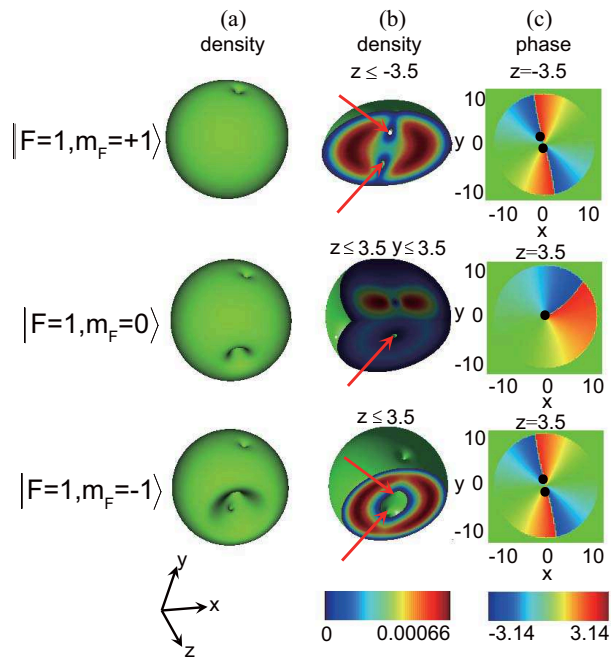}
\caption{(Color online). The monopole is metastable,
a doubly vortex line has split to two singly vortex
lines partly, as denoted by the red arrows, where
the SO coupling strength $\kappa=0$, the spin-dependent
interaction parameter $\lambda_{2}=-75$, the spin-independent
interaction parameter $\lambda_{0}=7500$, the strength of
the magnetic field gradient $B_{1}=0.6$, and the optical
trap frequencies $\omega_{r}=\omega_{z}=2\pi\times250$Hz.
(a) Isosurface of the particle densities. (b)
Segments of the isosurface of the particle densities.
(c) Phase distributions of slice planes, the black dots
indicate the locations of vortices. The unit of the length
is $\sqrt{\hbar/m\omega}$, the unit of the strength of the
magnetic field gradient is $\omega\hbar/(g_{F}\mu_{B}a_{h})$,
the unit of the SO coupling strength is $\sqrt{\hbar\omega/m}$,
and the scale for three dimension system is $34.2\mu m\times34.2\mu m\times34.2\mu m$.
} \label{figs2}
\end{figure}

\begin{figure}
\includegraphics[width= 0.48\textwidth]{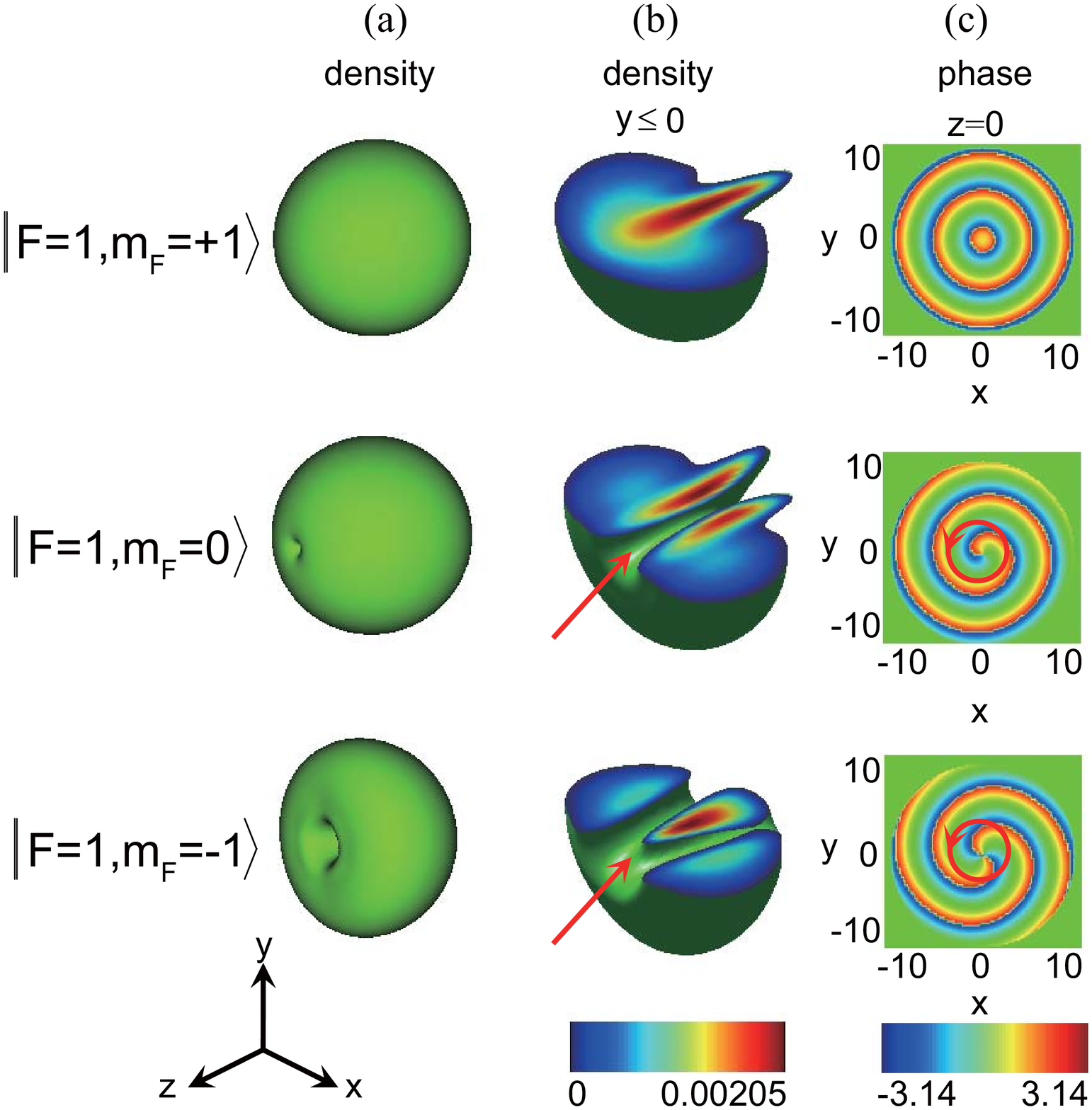}
\caption{(Color online). The M-MHV are obtained, which behave
as a soliton in the $m_{F}=+1$ component, a vortex
line with unit circulation in the $m_{F}=0$ component,
and a doubly quantized vortex line in the $m_{F}=-1$
component, where the SO coupling strength $\kappa=2$,
the spin-dependent interaction parameter $\lambda_{2}=-15$,
the spin-independent interaction parameter $\lambda_{0}=7500$,
the strength of the magnetic field gradient $B_{1}=0.6$, and
the optical trap frequencies $\omega_{r}=\omega_{z}=2\pi\times250$Hz.
(a) Isosurface of the particle densities. (b) Segments of the
isosurface of the particle densities ($y\leq0$), where the red
arrows indicate the location of the one nodal line (Dirac string).
(c) Phase distributions of the $z=0$ slice planes,
the circulations of the single vortex (the second line)
and the a doubly vortex (the third line) are the same,
which is denoted by the black circles. The unit of the
length is $\sqrt{\hbar/m\omega}$, the unit of the strength
of the magnetic field gradient is $\omega\hbar/(g_{F}\mu_{B}a_{h})$,
the unit of the SO coupling strength is $\sqrt{\hbar\omega/m}$, and
the scale for three dimension system is $34.2\mu m\times34.2\mu m\times34.2\mu m$.
} \label{figs3}
\end{figure}

\section*{The structures of the monopoles in the absence of the spin-orbit coupling}

Fig. 7 shows the results of the monopoles in the
absence of the SO coupling, where the dimensionless
SO coupling strength $\kappa=0$, the dimensionless
spin-dependent interaction parameter $\lambda_{2}=-75$,
the dimensionless spin-independent interaction
parameter $\lambda_{0}=7500$, the dimensionless
strength of the magnetic field gradient $B_{1}=0.6$,
and the optical trap $\omega_{r}=\omega_{z}=2\pi\times250$ Hz \cite{1}.
Note that a doubly quantized vortex line partly splits
into two singly quantized vortex lines in the $m_{F}=+1$
and $-1$ components [as highlighted by the red arrow in
Fig. 7(b)]. In Ref. \cite{9}, the external magnetic
field of the monopole creation is the combination of
the quadrupole field and the homogeneous bias field,
the vortex lines are located in the direction of the
bias field. However, in our case, a homogeneous bias
field is not considered in the experimental setup of
the monopole creation, which gives rise to the splitting
of the doubly quantized vortex line. The structures of
the vortex lines for the $m_{F}=+1$ and $-1$ components
are similar. The vortex lines for the $m_{F}=+1$
component locate in the part of $z<0$, the vortex
lines for the $m_{F}=-1$ component locate in the part
of $z>0$, and such the vortex lines extend outwords
along the $\pm y$ directions to the boundary of the
BECs. The phases at both sides
of the vortex line are the same. For the case of the
$m_{F}=0$ component, the two vortex lines in the $x=0$
plane cross each other and topologically invariant
winding number is 1. The phases of the vortex line
along the $z$ and $y$ axes are opposed, being similar
to the vortex-antivortex pair  \cite{10,11}.
In fact, the doubly quantized vortex will split into
two singly quantized vortices absolutely as the time
goes on \cite{12,13}. Therefore, the results
suggest that the monopole is metastable in the absence of the SO coupling.

\section*{The monopoles with the Mermin-Ho vortex}

In this section, the spin-dependent
interaction parameter $\lambda_{2}=-15$,
the SO coupling strength $\kappa=2$, the other
parameters are chosen as the same with those
given for Fig. 7. The M-MHV \cite{14} are obtained, as shown in
Fig. 8. The structure of the M-MHV represents a soliton in the $m_{F}=+1$ component,
a singly vortex line in the $m_{F}=0$ component and a doubly
quantized vortex line in the $m_{F}=-1$ component. The vortices
in the $m_{F}=0$ and $m_{F}=-1$ components have the same
phase winding direction, which is denoted by the black circles [Fig. 8(C)].
The phases at both sides of $x=0$ and $y=0$ slice planes are inverse.
Such the structure of the monopole also emerges in spinor BECs with a magnetic field \cite{9}. Our results show that
the M-MHV can exist in the spinor
BECs with non-Abelian gauge field.

\section*{The effect of the anisotropic optical trap on the monopoles}

We also investigate the monopole structures for an anisotropic
optical trap, where $\kappa=2$, $\lambda_{2}=-75$, $B_{1}=0.6$,
and $\lambda_{0}=7500$. For comparison, we first consider the
isotropic optical trap with $\omega_{r}=\omega_{z}=2\pi\times250$ Hz \cite{1},
as shown in Figs. 9(a1)-9(a3). The monopoles with the polar-core
vortex (M-PCV) are obtained. The monopoles behave as a singly vortex line
in the $m_{F}=+1$ component, a soliton in the $m_{F}=0$ component,
and a singly antivortex line in the $m_{F}=-1$ component. In Figs. 9(b1)-9(b3),
the trapping frequencies are given by $\omega_{r}\approx2\pi\times160$ Hz
and $\omega_{z}\approx2\pi\times220$ Hz \cite{9}, and the corresponding
the anisotropy parameters
related to the optical trap $\gamma_{r}=1$ and $\gamma_{z}=1.375$.
We can find that the M-PCV disappear,
and at the same time Dirac string embedded in the BECs
splits into two strings with a singly quantized vortex. In this case, our
results suggest that this anisotropic trapping potential leads to the
metastable monopole states. In fact, the M-PCV exist in the condensates when the anisotropy of the optical trap
is small, being less than the magnitude of $\gamma_{r}=1$ and $\gamma_{z}=1.375$.
In Figs. 9(c1)-9(c3), the trapping frequencies are given by
$\omega_{r}\approx2\pi\times160$ Hz and $\omega_{z}\approx2\pi\times320$ Hz, and the anisotropy parameters
related to the optical trap $\gamma_{r}=1$ and $\gamma_{z}=2$. We observe
that the M-PCV still exist in the
BECs. The results confirm that the M-PCV can occur in the presence of the oblate trap.

\begin{figure}
\includegraphics[width= 0.82\textwidth]{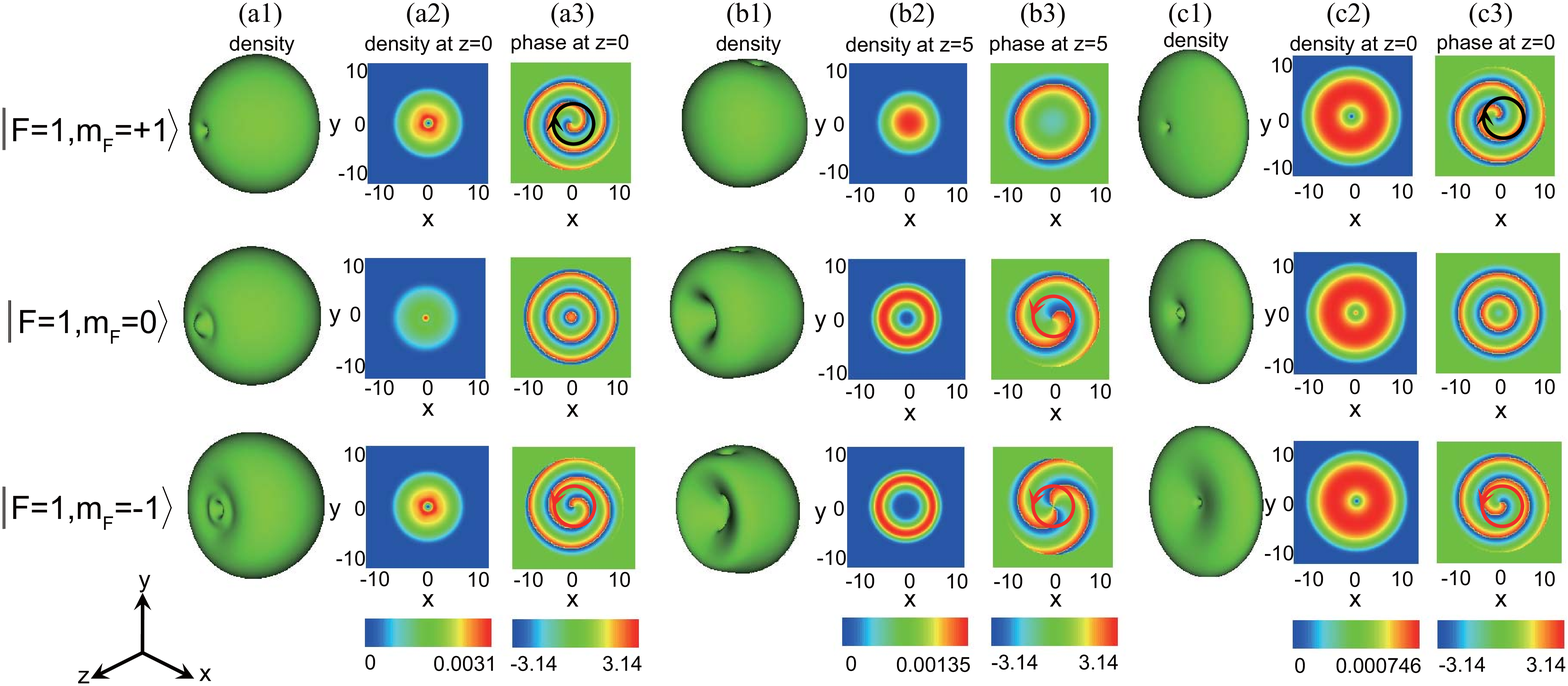}
\caption{(Color online). The monopole states for
the anisotropic trap as obtained by using the strength
of the magnetic field gradient $B_{1}=0.6$, the spin-independent
interaction parameter $\lambda_{0}=7500$, the spin-dependent
interaction parameter $\lambda_{2}=-75$ and the SO coupling
strength $\kappa=2$. (a1)-(a3) The optical trap frequencies
are given by $\omega_{r}=\omega_{z}=2\pi\times250$Hz, the
anisotropy parameters related to the optical trap $\gamma_{r}=\gamma_{z}=1$,
the M-PCV are obtained.
(b1)-(b3) The optical trap frequencies are given by
$\omega_{r}\approx2\pi\times160$Hz and $\omega_{z}\approx2\pi\times220$Hz,
the anisotropy parameters related to the optical trap $\gamma_{r}=1$
and $\gamma_{z}=1.375$, the M-PCV disappear.
(c1)-(c3) The optical trap frequencies are given by
$\omega_{r}\approx2\pi\times160$Hz and $\omega_{z}\approx2\pi\times320$Hz,
the anisotropy parameters related to the optical trap $\gamma_{r}=1$
and $\gamma_{z}=2$, the M-PCV remain exist
in the system. (a1), (b1) and (c1) show the isosurface of the particle
densities.  (a2), (b2) and (c2) display the particle densities of the slice
planes. (a3), (b3) and (c3) show the phase distributions of the slice planes.
The vortex and antivortex are denoted by the black circle of arrow and the
red circle of arrow. The unit of the length is $\sqrt{\hbar/m\omega}$, the
unit of the strength of the magnetic field gradient is $\omega\hbar/(g_{F}\mu_{B}a_{h})$,
the unit of the SO coupling strength is $\sqrt{\hbar\omega/m}$, and the scale for
three dimension system is $34.2\mu m\times34.2\mu m\times34.2\mu m$.
} \label{figs4}
\end{figure}

\begin{figure}
\includegraphics[width= 0.82\textwidth]{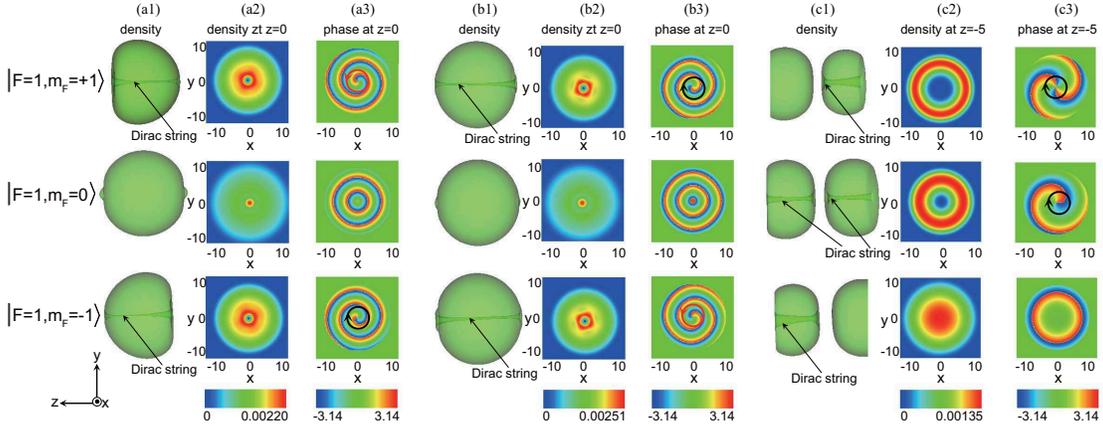}
\caption{(Color online). The monopole states for different quadrupole
field as obtained by using the optical trap frequencies $\omega_{r}=\omega_{z}=2\pi\times250$Hz,
where the spin-independent interaction parameter $\lambda_{0}=7500$,
the spin-dependent interaction parameter $\lambda_{2}=-75$, and
the SO coupling strength $\kappa=2$. (a1)-(a3) The strength
of the magnetic field gradient $B_{1}=-0.6$, the structure of
the antimonopole is as follows: a singly antivortex line in
 the $m_{F}=1$ component, a soliton in the $m_{F}=0$ component,
 and a singly vortex line in the $m_{F}=-1$ component.
 (b1)-(b3) The strength of the magnetic field gradient $B_{1}=0.2$,
 the structure of the monopole is as follows: a singly
 vortex line in the $m_{F}=1$ component, a soliton in
 the $m_{F}=0$ component, and a singly antivortex line
 in the $m_{F}=-1$ component. (c1)-(c3)
 The strength of the magnetic field gradient $B_{1}=3.8$,
 the M-PCV disappear.
 (a1), (b1) and (c1) show
 the isosurface of the particle densities, Dirac strings
 are denoted by the black arrows. (a2), (b2) and (c2)
 show the particle densities of the slice planes.
 (a3), (b3) and (c3) show
 the phase distributions of the slice planes. The vortex
 and antivortex are denoted by the black circle of arrow
 and the red circle of arrow. The unit of the length is
 $\sqrt{\hbar/m\omega}$, the unit of the strength of the
 magnetic field gradient is $\omega\hbar/(g_{F}\mu_{B}a_{h})$,
 the unit of the SO coupling strength is $\sqrt{\hbar\omega/m}$,
 and the scale for three dimension system is $34.2\mu m\times34.2\mu m\times34.2\mu m$.} \label{figs5}
\end{figure}

\begin{figure}
\includegraphics[width= 0.56\textwidth]{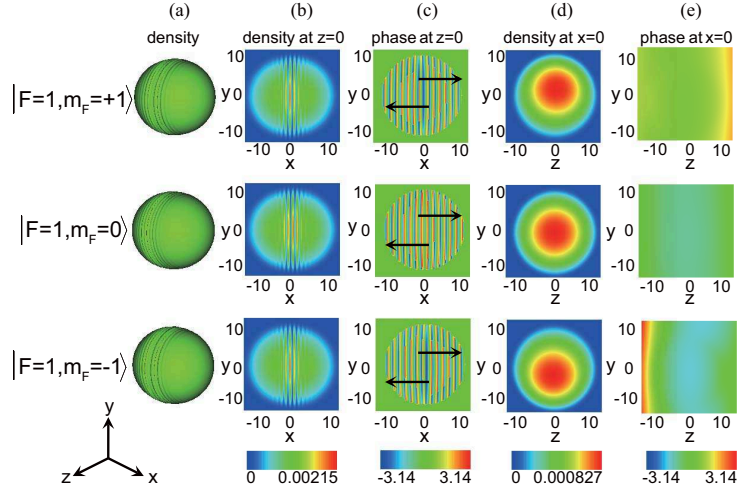}
\caption{(Color online). (a) Isosurface
of the particle densities of the three spin components,
where the anisotropic SO coupling $\kappa_{x}=8$ and
$\kappa_{y}=2$, the spin-dependent interaction parameter
$\lambda_{2}=-75$, the spin-independent interaction
parameter $\lambda_{0}=7500$, the strength of the
magnetic field gradient $B_{1}=0.6$, the optical trap
frequencies $\omega_{r}=\omega_{z}=2\pi\times250$Hz.
(b) and (c) Densities and phase
distributions of the stripe phase in the $x$-$y$ plane ($z=0$),
the phenomena in the $x$-$z$ plane are similar to those in
the $x$-$y$ plane. In (c), we can see that the
phases at both sides of the $z=0$ slice plane are inverse,
which is indicated by the black arrows. (d)
and (e) Densities and phase distributions
of the plane wave phase in the $y$-$z$ plane ($x=0$).
The unit of the length is $\sqrt{\hbar/m\omega}$, the
unit of the strength of the magnetic field gradient is
$\omega\hbar/(g_{F}\mu_{B}a_{h})$, the unit of the SO
coupling strength is $\sqrt{\hbar\omega/m}$, and the
scale for three dimension system is $34.2\mu m\times34.2\mu m\times34.2\mu m$.} \label{figs6}
\end{figure}

\begin{figure}
\includegraphics[width= 0.66\textwidth]{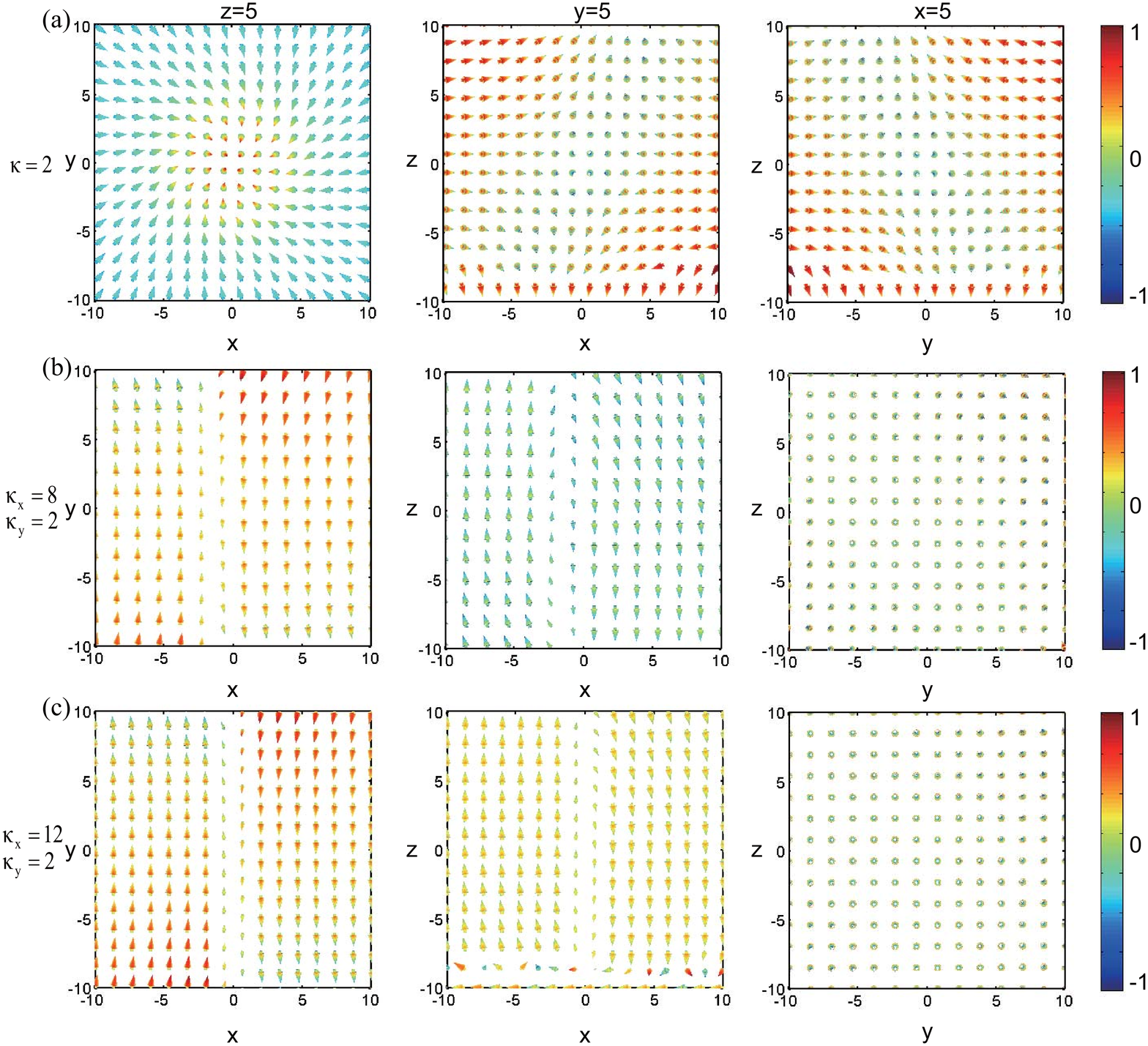}
\caption{(Color online). (a) When the SO
coupling $\kappa_{x}=\kappa_{y}=\kappa=2$, we obtain
the spin textures of the M-PCV in the slice planes of $z=5$, $y=5$ and $x=5$.
(b) When the SO coupling $\kappa_{x}=8$ and
$\kappa_{y}=2$, the spin textures show the stripe
distribution in the $z=5$ and $y=5$ planes and the
plane wave ferromagnetic distribution in the $x=5$
plane. (c) When the SO coupling $\kappa_{x}=12$
and $\kappa_{y}=2$, the spin textures are similar to
those in (b). For all figures, the strength
of the magnetic field gradient $B_{1}=0.6$, the
spin-dependent interaction parameter $\lambda_{2}=-75$,
the spin-independent interaction parameter $\lambda_{0}=7500$,
the optical trap frequencies $\omega_{r}=\omega_{z}=2\pi\times250$Hz,
the unit of the length is $\sqrt{\hbar/m\omega}$, the unit of the
strength of the magnetic field gradient is $\omega\hbar/(g_{F}\mu_{B}a_{h})$,
the unit of the SO coupling strength is $\sqrt{\hbar\omega/m}$, and
the scale for three dimension system is $34.2\mu m\times34.2\mu m\times34.2\mu m$.} \label{figs7}
\end{figure}

\begin{figure}
\includegraphics[width= 0.76\textwidth]{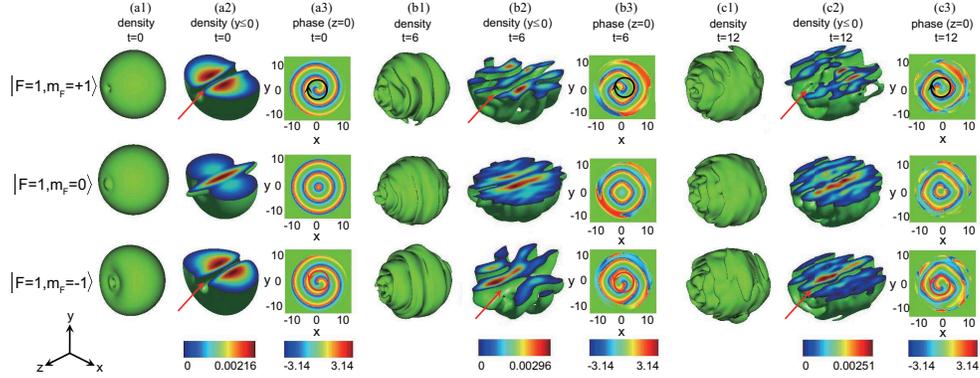}
\caption{(Color online). The dynamics of the M-PCV remain stable when the quadrupole field
is turned off, where the weak SO coupling strength $\kappa=2$,
the spin-dependent interaction parameter $\lambda_{2}=-75$,
spin-independent interaction parameter $\lambda_{0}=7500$,
the optical trap frequencies $\omega_{r}=\omega_{z}=2\pi\times250$Hz,
and the real time $t'=0.64t$ msec.
(a1)-(a3) t=0. (b1)-(b3) t=6.
(c1)-(c3) t=12. (a1), (b1)
and (c1) Isosurface of the particle densities.
(a2), (b2) and (c2) Particle
densities for $y\leq0$, where the red arrows indicate the location of the nodal lines
(Dirac strings). (a3), (b3) and (c3)
Phase distributions of the $z=0$ slice planes, the singly
vortex is denoted by the black circle of arrow and the singly
antivortex is denoted by the red circle of arrow. The unit of
the length is $\sqrt{\hbar/m\omega}$, the unit of the strength
of the magnetic field gradient is $\omega\hbar/(g_{F}\mu_{B}a_{h})$,
the unit of the SO coupling strength is $\sqrt{\hbar\omega/m}$, and
the scale for three dimension system is $34.2\mu m\times34.2\mu m\times34.2\mu m$.} \label{figs8}
\end{figure}

\begin{figure}
\includegraphics[width= 0.76\textwidth]{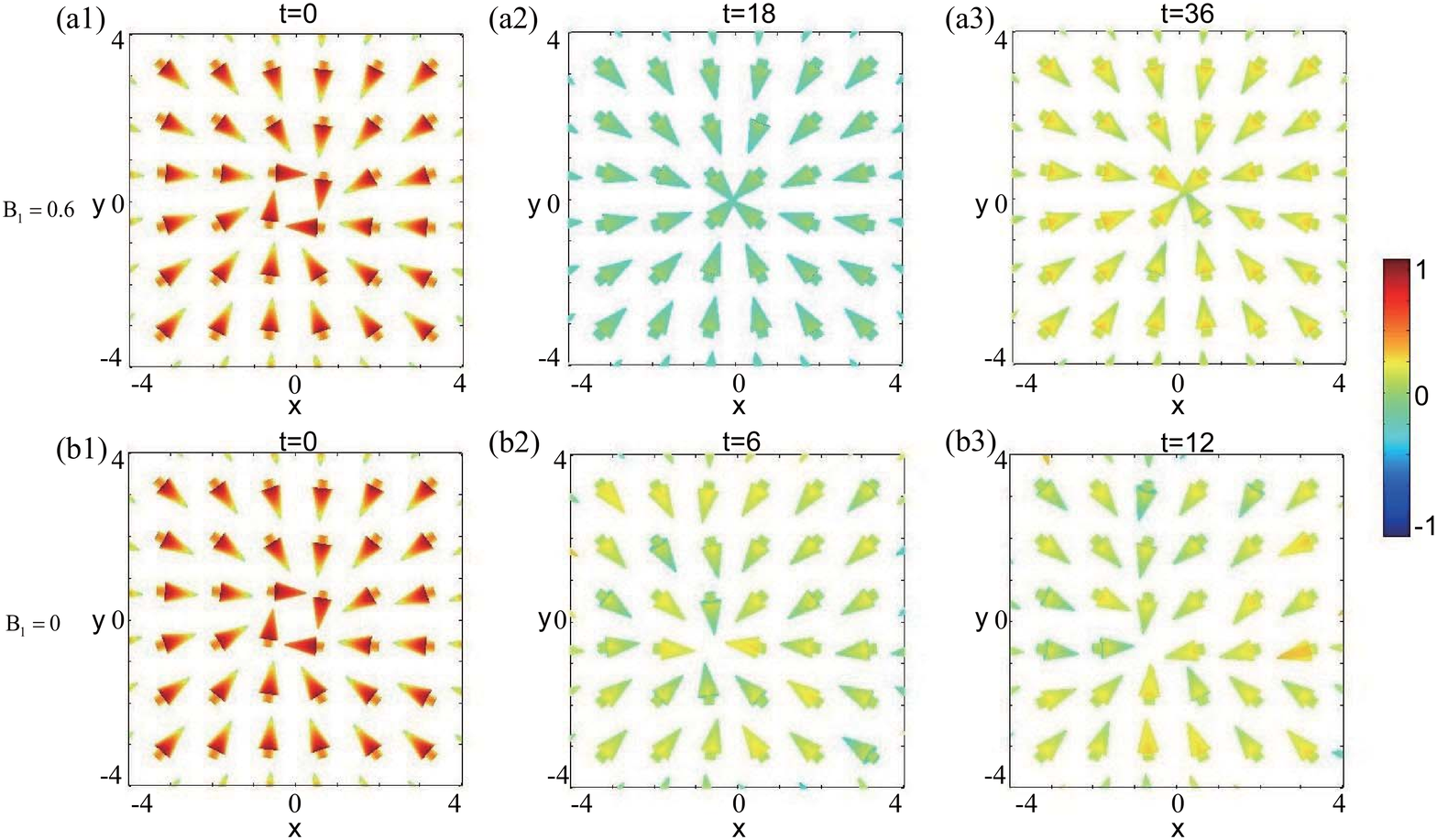}
\caption{(Color online). (a1)-(a3) Dynamic
evolution of the spin texture of the M-PCV in $x$-$y$ plane, where the strength of the magnetic
field gradient $B_{1}=0.6$, the weak SO coupling strength $\kappa=2$,
the spin-dependent interaction parameter $\lambda_{2}=-75$, spin-independent
interaction parameter $\lambda_{0}=7500$, the optical trap frequencies
$\omega_{r}=\omega_{z}=2\pi\times250$Hz, and the real time $t'=0.64t$ msec.
(a1) $t=0$. (a2) $t=18$. (a3) $t=36$.
(b1)-(b3) Dynamic evolution of the spin texture
of the M-PCV in $x$-$y$ plane, where the
strength of the magnetic field gradient $B_{1}=0$, other parameters
are the same as those in (a1) - (a3). (b1)
$t=0$. (b2) $t=6$. (b3) $t=12$. The unit of the
length is $\sqrt{\hbar/m\omega}$, the unit of the strength of the
magnetic field gradient is $\omega\hbar/(g_{F}\mu_{B}a_{h})$, the
unit of the SO coupling strength is $\sqrt{\hbar\omega/m}$, and
the scale for three dimension system is $34.2\mu m\times34.2\mu m\times34.2\mu m$.} \label{figs9}
\end{figure}

\section*{The effect of the quadrupole field gradient on the monopoles}
Because different external magnetic fields will change the
positions of Dirac strings and spin direction, so we investigate the influence of the quadrupole field
gradient on the monopoles, where $\kappa=2$, $\lambda_{2}=-75$,
$\lambda_{0}=7500$, and $\omega_{r}=\omega_{z}=2\pi\times250$ Hz.
In Figs. 10(a1)-10(a3), when the strength of the quadrupole field
gradient is negative, such as $B_{1}=-0.6$, the antimonopoles
emerges in the system. The structures of the antimonopoles in
terms of every spin components are as follows: a singly anti-vortex
line in the $m_{F}=+1$ component, a soliton in the $m_{F}=0$ component,
and a singly vortex line in the $m_{F}=-1$ component, where the phenomenon
is contrary, comparing with that of the monopoles with $B_{1}=0.6$.
In Figs. 10(b1)-10(b3), when $B_{1}$ is the small positive value such as
$B_{1}=0.2$, we find that the results in the case of weak quadrupole field
gradient are similar to those for $B_{1}=0.6$, which describes a singly
vortex line in the $m_{F}=1$ component, a soliton in the $m_{F}=0$ component,
and a singly antivortex line in the $m_{F}=-1$ component. In Figs. 10(c1)-10(c3),
when $B_{1}$ is the big positive value such as $B_{1}=3.8$, there is a large
difference, because the magnetic trap is strongest in $z$ direction. The result shows that the atomic cloud expands from
center to both sides in $z$ direction, which is caused that all
the atoms are difficult to be bounded in the central region of trap when
the strength of the magnetic field gradient increases. A vortex line is
terminated in the middle of the atomic cloud for the $m_{F}=+1$ and $m_{F}=-1$
components, and the phase winding of the vortex line is $4\pi$ between the
$m_{F}=+1$ and $m_{F}=-1$ component. In addition, for the $m_{F}=0$ component,
the vortex line locates in the positive and negative $z$ axis, the
corresponding phase winding number of the vortex line is 1.

\section*{Ground states for the anisotropic spin-orbit coupling}

In Fig. 11, we study the case of the anisotropic SO coupling,
where the strength is strong in the $x$ direction with $\kappa_{x}=8$,
but is weak in the $y$ direction with $\kappa_{y}=2$. The
spin-dependent interaction parameter $\lambda_{2}=-75$ and
the strength of the magnetic field gradient $B_{1}=0.6$.
The result shows that the monopole vanishes, which is caused
that the asymmetric spin-orbit coupling can remove the Dirac
string. Instead, the condensate splits into many segments
along the $x$ direction, representing a stripe phase along
the $x$ direction and a plane wave phase in $y$-$z$ plane.
The phases along the $x$ axis are inverse, which is indicated by the black arrows.

\section*{Spin textures for the anisotropic spin-orbit coupling}

Fig. 12 shows the spin textures of the spinor BECs with the anisotropic SO coupling. For comparison,
when the SO coupling is isotropic, such as $\kappa=2$,
the spin aligns with the radially inward hedgehog distribution
in the $x$-$y$ plane, while the spin textures in the $x$-$z$
and $y$-$z$ planes show the cross hyperbolic distribution [Fig. 12(a)].
When the SO coupling is anisotropic, such as $\kappa_{x}=8$ and
$\kappa_{y}=2$, the spin textures show the stripe distribution
in the $x$-$y$ and $x$-$z$ planes and the ferromagnetic
distribution in the $y$-$z$ plane [Fig. 12(b)]. Furthermore,
when the anisotropy of the SO coupling increases, such as
$\kappa_{x}=12$ and $\kappa_{y}=2$, we find that the spin
textures are similar to those in Fig. 12(b), which suggests
that the spin textures are almost not changed when the
strength of the anisotropic SO coupling increases [Fig. 12(c)].

\section*{The dynamic evolution of the monopoles in the absence of the quadrupole field}
In this section, we study the dynamics of the M-PCV when the quadrupole field is
turned off. The initial states of the monopoles are shown
in Figs. 13(a1)-13(a3). The monopoles are evolved up to
$12/\omega$ with a time step $10^{-4}/\omega$. During
the time evolution, the isosurface of density becomes
rough, see Fig. 13(b1) and Fig. 13(c1). However, the
Dirac strings still exist in the condensates, which
is denoted by the red arrows in Fig. 13(b2) and Fig. 13(c2).
Furthermore, as seen in the phase profile of the wave
function, a singly quantized vortex and antivortex do
not split, which is denoted by the black circle of
arrow and the red circle of arrow, as shown in Fig. 13(b3)
and Fig. 13(c3). Compared with the dynamics of the
monopoles in the previous work, in which the Dirac
strings are observed to expand at about $8/\omega$ \cite{3}.
In contrast, in our case, the Dirac strings is not observed
to expand until $12/\omega$, which suggests that the SO
coupling can protect such the monopole structures during the time evolution
in the absence of the external magnetic field.

\section*{The dynamic evolution of the spin texture}

Fig. 14 is the spin texture of the M-PCV at different times. Figs. 14(a1)-14(a3)
indicate the dynamic of the spin texture in the presence of
the quadrupole field. We note that the spin texture remains
the structure of the south magnetic pole during the time
evolution, where the spin aligns with the radially inward
hedgehog distribution in the $x$-$y$ plane. Furthermore,
when the quadrupole field is turned off, the spin texture
of the south magnetic pole remain immobile, see Figs. 14(b1)-14(b3).
We further suppose that the M-PCV are
stable by means of the dynamic evolution of the spin texture.

\end{widetext}

\end{document}